\documentstyle[12pt]{article}
\include{epsf}
\newlength{\extralineskip}
\newcommand{\junk}[1]{}

\newcommand{\Z}{Z\!\!\!Z}
\newcommand{\beq}{\begin{equation}}
\newcommand{\eeq}{\end{equation}}
\newcommand{\bd}{\begin{displaymath}}
\newcommand{\ed}{\end{displaymath}}
\def\bea{\begin{eqnarray}}
\def\eea{\end{eqnarray}}
\def\nn{\nonumber}
\def\ba{\beq\new\begin{array}{c}}
\def\ea{\end{array}\eeq}

\def\inbar{\,\vrule height1.5ex width.4pt depth0pt}
\def\IC{\relax\hbox{$\inbar\kern-.3em{\rm C}$}}
\def\IR{\relax{\rm I\kern-.18em R}}
\def\IN{\relax{\rm I\kern-.18em N}}

\makeatletter
\newdimen\normalarrayskip              
\newdimen\minarrayskip                 
\normalarrayskip\baselineskip
\minarrayskip\jot
\newif\ifold             \oldtrue            \def\new{\oldfalse}
\def\arraymode{\ifold\relax\else\displaystyle\fi} 
\def\@arrayskip{\ifold\baselineskip\z@\lineskip\z@
     \else
     \baselineskip\minarrayskip\lineskip2\minarrayskip\fi}
\def\@arrayclassz{\ifcase \@lastchclass \@acolampacol \or
\@ampacol \or \or \or \@addamp \or
   \@acolampacol \or \@firstampfalse \@acol \fi
\edef\@preamble{\@preamble
  \ifcase \@chnum
     \hfil$\relax\arraymode\@sharp$\hfil
     \or $\relax\arraymode\@sharp$\hfil
     \or \hfil$\relax\arraymode\@sharp$\fi}}
\def\@array[#1]#2{\setbox\@arstrutbox=\hbox{\vrule
     height\arraystretch \ht\strutbox
     depth\arraystretch \dp\strutbox
     width\z@}\@mkpream{#2}\edef\@preamble{\halign \noexpand\@halignto
\bgroup \tabskip\z@ \@arstrut \@preamble \tabskip\z@ \cr}%
\let\@startpbox\@@startpbox \let\@endpbox\@@endpbox
  \if #1t\vtop \else \if#1b\vbox \else \vcenter \fi\fi
  \bgroup \let\par\relax
  \let\@sharp##\let\protect\relax
  \@arrayskip\@preamble}

\newcommand{\mapto}{\to}
\begin{document}

\thispagestyle{empty}
\begin{flushright}
hep-th/9712173
\end{flushright}
\vspace*{10mm} \centerline{{\Large Topology and Duality in Abelian
Lattice Theories}\footnote{This work is supported in part by NSERC of
Canada, NATO CRG 970561 and the Fonds zur F\"orderung der
Wissenschaftlichen Forschung in \"Osterreich under project number
J01185-PHY.}} \vskip10mm \centerline{\bf C. R. Gattringer,
S. Jaimungal and G. W. Semenoff}
\vskip 10mm
\centerline{Department of Physics and
Astronomy}
\centerline{University of British Columbia}
\centerline{6224 Agricultural Road}
\centerline{Vancouver, British Columbia V6T 1Z1}
\vskip 10mm
\begin{abstract}
We show how to obtain the dual of any lattice model, with inhomogeneous
local interactions, based on an arbitrary Abelian group in any
dimension and on lattices with arbitrary topology. It is shown that
in general the dual theory contains disorder loops on the generators
of the cohomology group of a particular dimension. An explicit
construction for altering the statistical sum to obtain a self-dual
theory, when these obstructions exist, is also given. We discuss some
applications of these results, particularly the existence of
non-trivial self-dual 2-dimensional $\Z_N$ theories on the torus. In
addition, we explicitly construct the $n$-point functions of plaquette
variables for the $U(1)$ gauge theory on the 2-dimensional $g$-tori.
\end{abstract}
\vspace{70mm}

\setcounter{page}{0}
\setcounter{equation}{0} 
\newpage

It has been known for half a century \cite{KrWa41} that duality is a
powerful tool for analyzing statistical models \cite{Sa80}.  In recent
years, these ideas have been revived and have had impact in the study
of supersymmetric gauge theories and superstring theory.  The basic
idea is that, given a statistical, field theoretical or string model,
there may exist another representation of the model in which strong
and weak coupling limits are interchanged.  Most discussions of
duality in statistical systems apply to spaces which have trivial
topology.  However, there are many instances of physical interest,
such as regularized finite temperature gauge theory, where a
statistical system is defined on a space with non-trivial topology and
where some version of duality could be a useful tool.  In this Letter
we shall show that duality can indeed be implemented on a
topologically non-trivial space.  Our results apply to models with
degrees of freedom in Abelian groups which live on $k$-cells (sites,
links, plaquettes, etc.)  of a lattice that is the triangulation of a
$d$-dimensional smooth manifold.  We restrict our attention to models
with nearest-neighbor interactions.

The essence of our result is that Abelian duality on a lattice with
non-trivial homology requires the appearance of disorder defects on
cohomology cocycles of the lattice in either the original or the dual
theory.  These defects are similar to the familiar t'Hooft loops which
are used to characterize the phase structure of gauge theories. The
defects can be classified systematically and this allows one to
identify self-dual models on spaces with non-trivial topology.  For
example, given a model that is self-dual on an infinite space, one can
identify the modification of the statistical sum which makes the model
self-dual when some of the dimensions are compactified. We will
present as an example, a self dual Ising model and a self-dual $\Z_N$
spin model on the 2-torus.  Also, we explicitly compute the $n$-point
functions of plaquette variables for the $U(1)$ gauge theory on the
2-dimensional $g$-tori. 

To present our results in their most general form, we must introduce
some terminology.  We use the language of simplicial homology
\cite{Mu84,DrWa82}.  Consider a lattice $\Omega$ and associate to
every $k$-dimensional cell of the lattice an oriented generator
$c_{k}^{(i)}$ where $i$ indexes the various cells of dimension $k$.
These objects are used as generators of the $k$-chain group, denoted
by $C_k(\Omega,G)$, $$
\sum_{i=1}^{N_k} ~g_i ~c_{k}^{(i)} = g \in C_k(\Omega,G) \qquad,\qquad
g_i\in G $$ Here $G$ is an arbitrary Abelian group with group
multiplication implemented through addition and $N_k$ is the number of
$k$-cells in the lattice $\Omega$.  An element $g\in C_k(\Omega,G)$ is
called a $G$-valued $k$-chain or simply a $k$-chain.  Clearly
$C_k(\Omega,G) = \oplus_{i=1}^{N_k} G $.

Two homomorphisms, the boundary $\partial$ and the coboundary
$\delta$, define the chain complexes $(C_*(\Omega,G), \partial)$ and
$(C_*(\Omega,G), \delta)$ where $C_*(\Omega,G)\equiv
\oplus_{k=0}^{d} C_k(\Omega,G)$: \bea
0\stackrel{\partial_{d+1}}{\longrightarrow} C_{d}(\Omega,G)
\stackrel{\partial_{d}}{\longrightarrow}\dots
\stackrel{\partial_{k+2}}{\longrightarrow} C_{k+1}(\Omega,G)
\stackrel{\partial_{k+1}}{\longrightarrow}
C_k(\Omega,G)\stackrel{\partial_k}{\longrightarrow} C_{k-1}(\Omega,G)
\stackrel{\partial_{k-1}}{\longrightarrow} \dots
C_{0}(\Omega,G)
\stackrel{\partial_{0}}{\longrightarrow} 0 \nn\\ 0
\stackrel{\delta_{d}}{\longleftarrow} C_{d}(\Omega,G)
\stackrel{\delta_{d-1}}{\longleftarrow} \dots
\stackrel{\delta_{k+1}}{\longleftarrow} C_{k+1}(\Omega,G)
\stackrel{\delta_k}{\longleftarrow}
C_{k}(\Omega,G)\stackrel{\delta_{k-1}}{\longleftarrow}C_{k-1}(\Omega,G)\stackrel{\delta_{k-2}}{\longleftarrow}
\dots C_{0}(\Omega,G)
\stackrel{\delta_{-1}}{\longleftarrow} 0 \nn \eea 
These homomorphisms are defined by their
actions on the generators $c_k^{(i)}$ (we display the dimension
subscripts on $\partial_k$ and $\delta_k$ only when essential), \beq
\partial c_k^{(i)} =\sum_{j=1}^{N_{k-1}} [c_k^{(i)} : c_{k-1}^{(j)}]
~c_{k-1}^{(j)} ~~,~~ \delta c_k^{(i)} = \sum_{j=1}^{N_{k+1}} [
c_{k+1}^{(j)}:c_{k}^{(i)}] ~c_{k+1}^{(j)}
\label{cobd} \eeq where the incidence number is given by,
$$ [c_k^{(i)} : c_{k-1}^{(j)}] = \left \{
\begin{array}{ll} 
\pm 1 & \mbox{if the $j$'th $(k-1)$-cell is contained in the $i$'th
$k$-cell} \\ 0 & \mbox{otherwise}
\end{array}\right.$$ The plus or minus sign reflects the relative orientation of the
cells.  The boundary (co-boundary) chains and the exact (co-exact)
chains are defined as, 
$$
\begin{array}{lclclcl}
 B_k(\Omega, G) & = & {\rm Im}~\partial_{k+1} & , 
&B^k(\Omega,G) & = & {\rm Im}~\delta_{k-1} \cr
Z_k(\Omega, G) & = &{\rm ker}~\partial_k & , 
&Z^k(\Omega,G) &= &{\rm ker}~\delta_k 
\end{array}$$
These sets inherit their group structure from the chain complex.  The
boundary and co-boundary operators are nilpotent: $\partial\partial =
0$ and $\delta \delta =0$.  The quotient groups,
$$
H_k(\Omega, G) = Z_k(\Omega,G)/B_k(\Omega,G) \qquad , \qquad
H^k(\Omega, G) = Z^k(\Omega,G) / B^k(\Omega, G)
$$
are the homology and cohomology groups respectively. The elements of
the homology group are those chains that have zero boundary and are
themselves not the boundary of a higher dimensional chain, whilst the
elements of the cohomology group are those chains that have zero
coboundary and are themselves not the coboundary of a lower
dimensional chain.  The inner product on $C_k(\Omega,G)$ is defined by
its action on the generators, $$ \langle c_k^{(i)}, c_l^{(j)}\rangle
= \delta_{k,l}\delta^{i,j} $$ and is linear in both
arguments.  The coboundary operator and boundary operators are dual to
each other with respect to this inner product, 
\beq \langle \partial
c_k^{(i)}, c_{k-1}^{(j)} \rangle = \langle c_k^{(i)}, \delta
c_{k-1}^{(j)} \rangle \label{cbdual} 
\eeq 
Using this language, the models which we consider can be stated as
follows: the degrees of freedom take values in $G$ and are defined on
$(k-1)$-cells, and the Boltzmann weights are defined on the coboundary of a
$(k-1)$-chain., \beq Z = \sum_{g\in C_{k-1}(\Omega,G)} 
\prod_{i=1}^{N_{k}} B_i\left( \langle\delta g, c_k^{(i)} \rangle \right )
\label{model} \eeq Here, and throughout our discussion,
we allow the Boltzmann weights, $B_i$, to differ on each $k$-cell.
This generalization allows one to consider a large class of models,
for example some of our results are easily generalized to random bond
models.  Also, it enables one to use the partition function as a
generating function for certain correlators. Of particular interest is
the correlator of a disorder operator and an anti-disorder operator.
It is obtained from the partition function by shifting $\delta g$ by
a chain, $f$, which has support on an assembly of adjacent
$(d-k)$-dimensional objects denoted by $\Gamma$.  $\Gamma$ is
constructed by choosing two arbitrary $(k-1)$-cells on the dual
lattice and connecting them by an arbitrary path of $k$-cells. It is
then re-interpreted on the original lattice by taking the lattice-dual
of the path. The resulting object is $\Gamma$ and the chain $f$ is
explicitly given by the expression $a\sum_{i\in \Gamma} c_{d-k}^{(i)}$
for some $a\in G$. Clearly, there are as many disorder operators  as
there are elements in the group \cite{FrKa80}. 

A familiar example of a spin system is the Ising model which, in our
notation, has $G=\Z_2$ and $k=1$ so that for $g\in C_0(\Omega,G)$, 
$$ \langle\delta g, c_1^{(l)}\rangle = \sum_{i=1}^{N_0} g_i
~\langle\delta c_0^{(i)}, c_1^{(l)} \rangle = g_{i_1(l)}-g_{i_2(l)}
$$ where $i_1(l)$ and $i_2(l)$ are the end points of the link $l$.
The Boltzmann weight for aligned spins ( $g_{i_1(l)}-g_{i_2(l)}=0$ mod
2) is $e^{\beta}$ and for anti-alinged spins (
$g_{i_1(l)}-g_{i_2(l)}=1$ mod 2) it is $e^{-\beta}$ so that, 
$$ Z =
\sum_{\{g_i = 0,1\}}\prod_l \exp\left\{
\beta~\left(1-2\left(g_{i_1(l)}-g_{i_2(l)}~{\rm mod}~2\right) 
\right)
\right \} $$ More general models of this kind include Abelian gauge
theories, spin models in arbitrary dimensions, plaquette models,
etc. 

We shall expand the Boltzmann weight in (\ref{model}) in terms of the
characters, $\chi_R(g)$ of the irreducible representations $R\in G^*$,
\beq B_i(g)=\sum_{R\in G^*}b_i(R)\chi_R(g)\qquad,\qquad b_i(R) = \frac
1 {|G|} \sum_{h\in G} {\overline\chi_R(h)} ~B_i(h)
\label{CharExp} \eeq where $|G|$ is
the order of the group. In the case of a continuous group the
normalized sum over group elements is replaced by the Haar integration
measure.  Since $G$ is Abelian, $G^*$ inherits an Abelian group
structure, where the product (taken to be addition) is implemented via
the tensor product of representations of $G$. In particular: $aR+bS =
(\otimes_{i=1}^a R)~ \otimes~(\otimes_{i=1}^b S)$ for $a,b\in\Z$ and
$R,S\in G^*$.  This implies the following factorization properties for
the characters, \beq
\begin{array}{lcll}
\chi_R(h_1+h_2) & = & \chi_R(h_1)~\chi_R(h_2) & \hspace{1cm}R\in G^* \mbox{ and }h_1,h_2\in G \cr
\chi_R(a~h_1) & = & \chi_{a R}(h_1) = \chi_{\otimes_{i=1}^a R}(h_1) &\hspace{1cm}a \in \Z \label{chi2}
\end{array}
\eeq
We insert (\ref{CharExp}) into the partition function (\ref{model}) to obtain,
\bea Z &=& \sum_{g\in C_{k-1}(\Omega,G)}~ \prod_{i=1}^{N_{k}}~\sum_{r_i\in G^*} b_i\left(r_i\right) ~\chi_{r_i} \left(\langle \delta g,
c_k^{(i)} \rangle \right )\nn\\
&=&\sum_{g\in C_{k-1}(\Omega,G)}~ \sum_{r\in
C_k(\Omega,G^*)}~\prod_{i=1}^{N_{k}} \left\{b_i\left(\langle r,c_k^{(i)}
\rangle\right) ~\chi_{\langle r,c_k^{(i)}\rangle} \left(\langle \delta g,
c_k^{(i)} \rangle \right ) \right\} \label{step1} \eea 
Here, the
product over $k$-cells has been interchanged with the sum over group
representations.  Every $k$-cell has been associated with a
representation $r_i\in G^*$ and this information is encoded in the
$G^*$ valued $k$-chain, $r=\sum_{i=1}^{N_k} r_i c_k^{(i)}$.  Applying
the factorization properties (\ref{chi2}) to the product of characters
in (\ref{step1}) one finds, 
\bea \prod_{i=1}^{N_{k}}\chi_{\langle
r,c_k^{(i)} \rangle } \left(\langle \delta g, c_k^{(i)} \rangle
\right ) &=& \prod_{i=1}^{N_{k}}\chi_{\langle r,c_k^{(i)}\rangle}
\left(\sum_{j=1}^{N_{k-1}} g_j ~[c_k^{(i)}:c_{k-1}^{(j)}] \right ) =
\prod_{i=1}^{N_{k}}\prod_{j=1}^{N_{k-1}} \chi_{\langle
r,c_k^{(i)}\rangle} \left( g_j ~[c_k^{(i)}:c_{k-1}^{(j)}] \right )
 \nn \\
&=& \prod_{j=1}^{N_{k-1}} \chi_{\sum_{i=1}^{N_k} \langle
r,c_k^{(i)}\rangle[c_k^{(i)}:c_{k-1}^{(j)} ]} \left( g_j \right )
\label{charfac} \eea 
The representations in the subscript of $\chi$ can be
rewritten in a more transparent manner by noticing the following,
$$
\sum_{i=1}^{N_k} \langle r,c_k^{(i)}\rangle~[c_k^{(i)}:c_{k-1}^{(j)} ]
= \langle r , \sum_{i=1}^{N_k} [c_k^{(i)}:c_{k-1}^{(j)} ] c_k^{(i)}
\rangle = \langle r , \delta c_{k-1}^{(j)} \rangle = \langle \partial r,
c_{k-1}^{(j)} \rangle 
$$
which follows directly from linearity of the inner product and the
definition of the coboundary operator (\ref{cobd}).  It is now possible
to sum expression (\ref{charfac}) over $g$, \beq \sum_{g\in
C_{k-1}(\Omega,G)}~\prod_{j=1}^{N_{k-1}}\chi_{\langle
\partial r,c_{k-1}^{(j)}\rangle} \left(g_j\right
) = \prod_{j=1}^{N_{k-1}} \sum_{g_j\in G}\chi_{\langle \partial r,
c_{k-1}^{(j)} \rangle } \left(g_j \right) = |G|^{N_{k-1}}
\prod_{j=1}^{N_{k-1}} ~\delta\left(\langle \partial r, c_{k-1}^{(j)}
\rangle,0\right)
\label{rcycle}
\eeq The last equality follows due to the orthogonality of the
characters, $ |G|^{-1} \sum_{g\in G} ~\chi_R ( g) ~\overline\chi_S(g)
= \delta(R,S)$ for $ R,S \in G^*$. The partition function now depends
on $k$-chains with constraints, $$ Z = {|G|}^{N_{k-1}}\sum_{r\in
C_k(\Omega,G^*)}~\prod_{i=1}^{N_k} b_i\left(\langle r ,
c_k^{(i)}\rangle \right) ~\prod_{j=1}^{N_{k-1}} ~\delta\left(\langle
\partial r, c_{k-1}^{(j)} \rangle,0\right) $$ However, the
constraints simply force $r$ to be an exact chain ($\partial
r=0$). This implies that $r$ is the sum of a boundary chain and
representatives of elements of the homology group under inclusion into
the chain group, \beq r = \partial r' + \sum_{a=1}^{A_k} h_a \gamma_a
\label{cycle} \; \; \; , \; \; \; r' \in C_{k+1}(\Omega, G^*) \eeq
where $\{\gamma_a:a=1,\dots,A_k\}$ are the generators of $H_k(\Omega,
G^*)\cong \oplus_{a=1}^ {A_k} H_{k,a}(\Omega,G^*)$ and $h_a\in
H_{k,a}(\Omega,G^*)\hookrightarrow G^*$. We will use $h\in \widetilde
H_k \left( \Omega,G^* \right)$ to denote this inclusion, which is
necessary so that addition in the argument of the Boltzmann weight is
group multiplication in $G^*$ and not in
$H_{k,a}\left(\Omega,G^*\right)$.  Removing the constraints and
inserting this form for $r$ into the partition function one finds,
\beq Z ={|G|}^{N_{k-1}-d_{k+1}}\sum_{h\in \widetilde H_k(\Omega,
G^*)}~\sum_{r \in C_{k+1}(\Omega,G^*)}~\prod_{i=1}^{N_k}~b_i\left(
\langle h+\partial r, c_k^{(i)}\rangle \right) \label{dual1} \eeq Here
we have dropped the prime on the dummy index $r'$, and $d_k\equiv
\mbox{ dim ker } \partial_{k}$ which is required to prevent overcounting.

The final step in the duality transformation is to re-interpret
$c_k^{(i)}$ as elements on the dual lattice: so that we make the
association $c^{*(i)}_{d-k} \leftrightarrow c_k^{(i)}$ where
$c^{*(i)}_k$ live on the dual lattice $\Omega^*$. Under this
association $\langle \partial c_k^{(i)},c_l^{(j)} \rangle =
\langle\delta c_{d-k}^{*(i)},c_{d-l}^{*(j)}\rangle$ follows
naturally. With this reinterpretation the dual partition function is
given by the compact expression, \beq
Z={|G|}^{N^*_{d-k+1}-D^*_{d-k-1}}\sum_{h\in H^{d-k}(\Omega^*, G^*)}~\sum_{r
\in C_{d-k-1}(\Omega^*,G^*)}~ \prod_{i=1}^{N^*_{d-k}}~ b_i\left(
\langle h+\delta r, c_{d-k}^{*(i)}\rangle\right)\label{dualZ} \eeq
where $D^*_{d-k}=d_k$ and is the dimension of the kernel of
$\delta_{d-k}$ on the dual lattice.  Notice that $h$ is now an element in the $(d-k)$'th
cohomology group of the dual lattice, which is isomorphic to the
$k$'th homology group of the lattice.

We now see that, in general, the dual theory has additional disorder
terms corresponding to the generators of the cohomology group of the
dual lattice.  By comparing (\ref{dualZ}) and (\ref{model}), it is
clear that, in order for (\ref{model}) to be self-dual,
$H_k(\Omega,G^*)$ has to be trivial.  Of course, in addition, the
lattice must be self-dual, $k$ must equal $d-k$ so that the original
and dual degrees of freedom live on the same type of cells and $G$
must be isomorphic to $G^*$.\footnote{$G\cong G^*$ for any finite
Abelian group.  An example of a group where this does not hold is
$U(1)$ where $U(1)^*\cong \Z$.}  We interpret this duality in the
general sense where the Boltzmann factors $B_i(g)$ for each group
element $g$ can be regarded as independent coupling constants and the
dual Boltzmann factors $b_i(r)$ for each $r\in G^*\cong G$ are the
images of these constants under the duality transformation. In the
normal sense of duality one requires a more specific manner in which
$B(g)$ transforms into $b(r)$, in particular, one requires these to
have the same functional form.

\begin{figure}[htbp]
\epsfysize=2.5in
\hspace*{43mm}
\epsfbox[0 0 353 271] {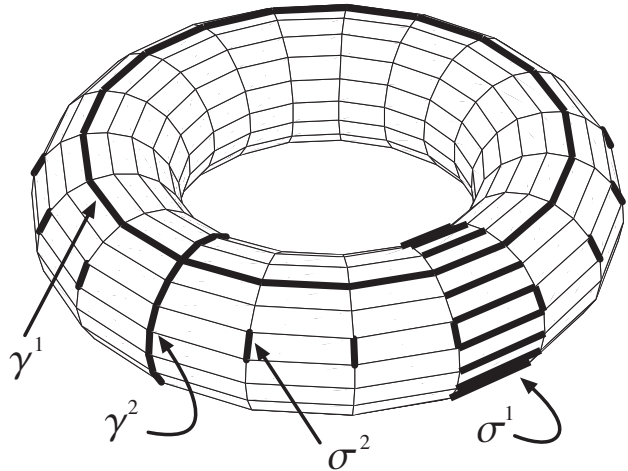}
\caption{This diagram depicts the cycles ($\gamma_i$) and cocycles
($\sigma^i$) of dimension 1 on the torus.\label{torusfig}}
\end{figure}

Of course, the Ising model on an infinite plane satisfies all the
requirements of self-duality. However, suppose one compactifies both
directions so that the model now lives on a torus. Topology is the
only obstruction to that model being self-dual. To see this explicitly
we write down the partition function for the dual model.  Firstly note
that the generators of $H^1(\Omega,G)$ are the cocycles $h^1 =
\sum_{l\in \sigma^1} c_1^{(l)}$ and $h^2=\sum_{l\in\sigma^2}
c_1^{(l)}$, where the sums are over all links in $\sigma^i$ shown in
figure 1. These cocycles are the duals (in the sense of
the inner product) to the cycles $\gamma_i$ which generate the
homology group. Using (\ref{dualZ}), the dual of the Ising model on
the torus has partition function, \bea Z&\propto& \sum_{h_1,h_2 = 0,1}
\sum_{\{r_i = 0,1\}} \exp \left\{ \beta^*\left[\sum_{l\notin \sigma^1
\cup \sigma^2} \left(1-2\left(r_{i_1(l)}-r_{i_2(l)}~{\rm mod
}~2\right)\right)\nn\right.\right.\\ &+&
\left.\left.\!\!\!\!\!\!\!\sum_{l\in\sigma^1}\!\!
\left(1-2\left(r_{i_1(l)}-r_{i_2(l)}+h_1~{\rm mod }~2\right)\right)+
\sum_{l\in\sigma^2}\!\!\left(1-2\left(r_{i_1(l)}-r_{i_2(l)}+h_2~{\rm
mod }~2\right)\right)\right]\right\} \nn \eea Here $\beta^*=-\frac12
\ln\tanh\beta$ is the well known dual coupling constant, also notice
that $\sigma^1 \cap \sigma^2=\emptyset$ thus all links are accounted
for in the Boltzmann weight.  The dual partition function is then the
sum of four copies of the Ising models in which the couplings along
$\sigma^i$ are taken to be $(\beta^*,\beta^*)$, $(-\beta^*,\beta^*)$,
$(\beta^*,-\beta^*)$ and $(-\beta^*,-\beta^*)$.  The extra terms have
the interpretation of disorder defects \cite{CeKa71,FrKa80}.

We have shown that the dual theory on a lattice with non-trivial
cohomology contains extra topological terms. In order to formulate
self-dual models in these cases it is necessary to begin with a theory
that contains a subset of the cohomology generators.  We shall see
that under duality these extra terms cancel some of the topological
terms that would appear in the dual theory. Such models have partition
functions given by, \beq Z = \sum_{h\in \widetilde
H^k_{(A)}(\Omega,G)} ~\sum_{g\in C_{k-1}(\Omega,G)}
\prod_{i=1}^{N_{k}} B_i\left(\left\langle (\delta g+h), c_k^{(i)}
\right\rangle \right ) \label{modmodel} \eeq where $\widetilde
H^k_{(A)}(\Omega,G)$ is a subgroup of $\widetilde H^k(\Omega,G)$
generated by a subset of the generators of the cohomology group,
$\{\sigma^a:a\in A\subseteq\{1,\dots,A_k\}\}$.  An element
$h\in\widetilde H^k_{(A)}(\Omega,G)$ is written as, $h=\sum_{a\in A}
h^a \sigma^a$ with $h^a\in H^{k,a}(\Omega,G)\hookrightarrow G$. As
before, the inclusion is necessary so that addition in the argument of
the Boltzmann weight is group multiplication in $G$ rather than in
$H^{k,a}(\Omega,G)$.

Once again the first step in the duality transformation is to perform a
character expansion of the Boltzmann weights, $$ Z = \sum_{r\in
C_k(\Omega,G^*)}~\sum_{h\in\widetilde H^k_{(A)}(\Omega,G)} ~\sum_{g\in
C_{k-1}(\Omega,G)} \prod_{i=1}^{N_{k}} b_i(\langle r,
c_k^{(i)}\rangle) ~\chi_{\langle r, c_k^{(i)}\rangle} \left (
\left\langle (\delta g+h), c_k^{(i)} \right\rangle \right ) $$ In
the above we have encoded the representations that live on $k$-cells
in the $k$-chain $r$, and introduced the character coefficients
$b_i\left(\langle r, c_k^{(i)}\rangle\right)$ as in equation
(\ref{CharExp}). The properties of the characters (\ref{chi2})
factorize the partition function, $$
 Z =\sum_{r\in
C_k(\Omega,G^*)}\prod_{i=1}^{N_{k}} b_i(\langle r, c_k^{(i)}\rangle)
\sum_{g\in C_{k-1}(\Omega,G)} \prod_{j=1}^{N_k} \chi_{\langle r,
c_k^{(j)}\rangle} \left ( \left\langle \delta g, c_k^{(j)}
\right\rangle \right ) \sum_{h\in\widetilde H^k_{(A)}(\Omega,G)}
\prod_{l=1}^{N_k}\chi_{\langle r, c_k^{(l)}\rangle} \left(
\left\langle h, c_k^{(l)} \right\rangle \right )  $$ The sum
over $g$ was performed previously and produced a delta function
forcing $r$ to be an exact chain (see (\ref{charfac}) and
(\ref{rcycle})). The sum over the cohomology elements will force
additional constraints on the representations.  Using the
factorization properties of the characters and the explicit
representation $h=\sum_{a\in A} h^a\sigma^a$ one finds, $$
\chi_{\langle r, c_k^{(i)}\rangle} \left( \left\langle h, c_k^{(i)}
\right\rangle \right ) = \prod_{a\in A} \chi_{\langle r,
c_k^{(i)}\rangle} \left( h^a \langle \sigma^a, c_k^{(i)} \rangle
\right ) = \prod_{a\in A} \chi_{\langle r, c_k^{(i)} \rangle\langle
\sigma^a, c_k^{(i)} \rangle} \left( h^a \right ) $$ Performing the
sum over $H^{k,a}$ and product over $k$-cells yields, \bea \sum_{h\in
\widetilde H^k_{(A)}(\Omega,G)}
&&\!\!\!\!\!\!\!\!\prod_{i=1}^{N_k}\chi_{\langle r, c_k^{(i)}\rangle}
\left( \left\langle h, c_k^{(i)} \right\rangle \right ) = \prod_{a\in
A} \sum_{h^a \in \widetilde H^{k,a}(\Omega, G)} \prod_{i_a=1}^{N_k}
\chi_{\langle r, c_k^{(i_a)} \rangle\langle \sigma^a, c_k^{(i_a)}
\rangle} \left( h^a \right ) \nn\\ = &&\!\!\!\!\!\!\!\prod_{a\in
A}~\sum_{h^a \in \widetilde H^{k,a}(\Omega, G)}
\chi_{\sum_{i_a=1}^{N_k}\langle r, c_k^{(i_a)} \rangle\langle
\sigma^a, c_k^{(i_a)} \rangle} \left( h^a \right ) = \prod_{a\in
A}~\sum_{h^a \in \widetilde H^{k,a}(\Omega, G)} \chi_{\langle r ,
\sigma^a \rangle} \left ( h^a \right ) \label{hsum} \eea Now
interpreting the inner product $\langle r , \sigma^a \rangle$ as a
representation of $H^{k,a}(\Omega,G)$, which can be done since
$H^{k,a}(\Omega,G)$ is a quotient subgroup of $G$, and performing the
sum over $h^a$ forces these representations to be the trivial
ones. Then (\ref{hsum}) reduces to a product of delta functions,
$\prod_{a\in A} |H^{k,a}(\Omega,G)| \delta\left(\langle r , \sigma^a
\rangle,0\right)$.

Putting this information together we obtain the very compact
expression for the partition function, $$ Z = |G|^{N_{k-1}-d_{k+1}}
\prod_{a'\in A} |H^{k,a'}(\Omega,G)|\sum_{r\in C_k(\Omega,G^*)}
\left(\prod_{i=1}^{N_{k}} b_i(\langle r, c_k^{(i)}\rangle)\right)
~\delta\left(\partial r,0\right) ~\prod_{a\in A} \delta\left(\langle r
, \sigma^a \rangle,0\right) $$ It is possible to solve the
constraints and remove the delta functions. Since $r$ is forced to be
exact, take it to be of the form (\ref{cycle}).  The other constraints
then allows one to solve for some of the coefficients $h_b$. On
lattices with no boundaries or torsion there exists an isomorphism
between cocycles and cycles induced by the inner product. The
generators can then be paired in the following way: $\langle \gamma_a
, \sigma^a \rangle = 1$ for $a=1,\dots,A_k$ (the inner product is
unity since one can always choose the canonical generators which
intersect on only one cell; compare figure 1.)
Restricting ourselves to lattices that have these
properties and solving the other constraints leads to, $$\langle
\partial r', \sigma^a \rangle + \sum_{b=1}^{A_k} h_b \langle \gamma_b,
\sigma^a\rangle =0 \quad \Rightarrow \quad h_a = -\langle \partial r',
\sigma^a \rangle = -\langle r', \delta \sigma^a\rangle = 0 \quad
\mbox{ for } a \in A $$ the last equality follows since
$\{\sigma^a\}$ are cocycles. Thus we see that the extra constraints
force the coefficients of the cycles "dual" to the cocycles to
vanish. Inserting these constraints into the partition function one
obtains, $$ Z = |G|^{N_{k-1}-d_{k+1}} \prod_{a'\in A}
|H^{k,a'}(\Omega,G)|\sum_{h\in\widetilde H_k^{(\overline
A)}(\Omega,G^*)}~ \sum_{r\in C_{k+1}(\Omega,G^*)}\prod_{i=1}^{N_{k}}
b_i\left( \left\langle(\partial r +h), c_{k}^{(i)} \right\rangle
\right) $$ In the above $\overline A$ denotes the compliment of the
set $A$, so that the homology subgroup is generated by those cycles
that have no counterpart in the modified model (\ref{modmodel}). Now
interpreting this partition function on the dual lattice by making the
associations mentioned earlier we find, $$ Z = |G|^{N^*_{d-k+1}-D^*_{d-k-1}}
\prod_{a'\in A} |H_{d-k,a'}(\Omega,G)|\sum_{h\in\widetilde
H^{d-k}_{(\overline A)}(\Omega^*,G^*)}~ \sum_{r\in
C_{d-k-1}(\Omega^*,G^*)}\prod_{i=1}^{N^*_{d-k}} b_i\left(
\left\langle(\delta r +h), c_{d-k}^{*(i)} \right\rangle \right) $$
The generators of $\widetilde H^{d-k}_{(\overline A)}(\Omega^*,G^*)$
are the cocycles on the dual lattice that are associated with the set
of cycles $\{ \gamma_b : b\in \overline A\}$ on the original
lattice. In the case of the Ising model with one cocycle,
$\{\sigma^1\}$, added in at the start the constraints eliminate
$\{\gamma_1\}$, so that $\{\gamma_2\}$ survives. The dual theory then
has an extra term supported on the lattice-dual to $\{\gamma_1\}$,
denoted by $\{\gamma^{*1}\}$ however, that cocycle is $\{\sigma^1\}$.

In order to illustrate the physical content of the ideas in this
paper, we discuss the $\Z_N$ model on the torus with topological terms
in some detail. We will denote elements of $\Z_N$ by $n$, the
representations of $\Z_N$ are isomorphic to $\Z_N$ and as such we
label them by the integers $r=0,1,\dots,N-1$. The characters are given
by $\chi_r(n)= \exp( i rn 2\pi/N)$.

We discuss the most general case on the torus by choosing one of the
subsets of the generators of the first Homology group: ${\bf s_1} =
\{\emptyset\}$, ${\bf s_2} = \{\sigma^1\}$, ${\bf s_3} = \{\sigma^2\}$
or ${\bf s_4} = \{ \sigma^1,\sigma^2\}$ (compare figure 1.) For a given
choice of ${\bf s_a}$ define the model by the following partition
function, $$ Z({\bf s_a}, \beta) =  \sum_{\{n_i=0\}}^{N-1}
\prod_{l\notin {\bf s_a}} B\left(n_{i_1(l)} - n_{i_2(l)}~{\rm
mod}N\right) \prod_{\sigma \in {\bf s_a}} \sum_{\{h_\sigma =0\}}^{N-1}
\prod_{l\in \sigma} B\left(n_{i_1(l)} - n_{i_2(l)} + h_\sigma~{\rm
mod}N\right) $$ for $a=1,2,3$ or 4. In Villain form the Boltzmann
factor is given by (equal on all links) \beq B(n) \; = \; \sum_{j \in
\Z} \exp \left\{ -\frac{\beta}{2} \Big( \frac n N - j \Big)^2 \right\}
\; .
\label{boltzfact}
\eeq
The coefficients $b(r)$ of the character expansion are given by,
\beq
b(r) = \frac{1}{N} \sum_{n=0}^{N-1} \sum_{j \in {\Z}} 
e^{ -\frac{\beta}{2} (
\frac nN - j )^2 } e^{-i \frac{2\pi}{N}r n}
= \frac{1}{N} \sum_{n \in {\Z}} 
e^{ -\frac{\beta}{2N^2}  n^2 } e^{-i \frac{2\pi}{N}rn}
 = \sqrt{ \frac{2 \pi}{ \beta}} \sum_{m \in {\Z}}
e^{-\frac{\beta^*}{2} (\frac rN - m)^2 } 
\label{boltzdual}
\eeq Here $\beta^* = (2\pi)^2N^2\beta^{-1}$ is the dual coupling
constant.  Thus we see that $b(r)$ has the same functional form as
$B(n)$.  By applying the above discussed rules on the cancellation of
the cycles we find the following transformation properties of $Z({\bf
s_a},\beta)$ under the duality transformation, $$ Z({\bf s_a}, \beta)
\; = \; \left(\frac{4 \pi}{\beta}
\right)^{N_0} \sum_{b=1}^4 M_{a,b} \; Z({\bf s_b}, \beta^*) $$ where the
transformation matrix $M_{a,b}$ is given by $$ M = \frac{1}{2}
\left( \begin{array}{cccc} 0 & 0 & 0 & 1 \\ 0 & 2 & 0 & 0 \\ 0 & 0 & 2
& 0 \\ 4 & 0 & 0 & 0
\end{array} \right) 
$$
The matrix $M$ obeys $M^2 = 1$ as should be since applying the duality
transformation twice gives the original model. A similar structure for the case of the homogeneous
Ising model in a different approach was discussed in \cite{BuSh96}.
The appearance of cocycles in duality transformations was also
appreciated by Rakowski and Sen \cite{RaSe97}, however they did not discuss
self-duality.

Self-dual and anti-self-dual models can be constructed using the
eigenvectors of $M$. Since $M^2=1$, the eigenvalues are either $+1$ or
$-1$.  In particular three eigenvectors have eigenvalue 1 giving rise
to the self dual models $Z({\bf s_2}, \beta), ~Z({\bf s_3}, \beta)$ and
$Z({\bf s_1}, \beta) + 2Z({\bf s_4}, \beta)$.  The fourth eigenvector
with eigenvalue $-1$ corresponds to the anti-self-dual model $-Z({\bf
s_1}, \beta) + 2 Z({\bf s_4}, \beta)$.

This strategy of choosing a complete set of cycles, computing the
corresponding dual theories and finding eigenvectors of the
transformation matrix $M$ is a method for finding self-dual theories
which can easily be implemented for other models and in higher
dimensions. In general, $M$ contains a single non-zero entry in each
row.  Since M is non-singular (in fact $M^2=1$), this implies that
every column also has only one non-zero entry.

As a second example, we show how one can use the inhomogeneous form of
the partition function to compute correlators.  In particular we
compute some correlators of lattice gauge theories on 2-dimensional
$g$-tori. In this case we choose $k=2$ so that the degrees of freedom
live on links of the lattice, and leave the group $G$ arbitrary for
the moment. The partition function is, $$ Z = \sum_{\{g_l\in G\}}
\prod_{p\in \Omega} B_p \left( \prod_{l\in p} g_l\right) $$ Here $l$
and $p$ labels the links and plaquettes of the lattice respectfully,
and we have reverted to the multiplicative notation for group
multiplication.  Since our formalism allows for inhomogeneous
Boltzmann weights it is possible to compute the expectation value of
arbitrary $n$-point functions of plaquette variables in arbitrary
representations.  This includes eg.~the filled Wilson loop. We alter
the weights of the Boltzmann factors for the plaquettes in the support
$\Gamma$ of our $n$-point function as follows: $$ B_p(g) \mapto
\chi_{S_p}(g) B_p(g) = \sum_{R
\in G^*} b_p(R) \chi_R(g)\chi_{S_p}(g) = \sum_{R\in G^*} b_p(R-S_p)
\chi_R(g) \quad,\quad \mbox{for all } p\in\Gamma $$ This introduces a
plaquette variable in representation $S_p$ for all plaquettes $p$ in
$\Gamma$. The dual partition function is rather trivial since there
are no cells of dimension negative one, and $H_0(\Omega,G^*)\cong
G^*$. Using this fact, the $n$-point function is given by, $$
\left\langle \prod_{p \in \Gamma} \chi_{S_p}(g) \right\rangle =  
\frac{\sum_{h\in G^*} \prod_{i\notin \Gamma^*} b_i(h) \prod_{i\in\Gamma} b_i(h-S^*_i)}{ \sum_{h\in G^*} \prod_{i\in\Omega} b_i(h)} 
$$ here $i$ labels the sites of the dual lattice, $\Gamma^*$ is the
lattice-dual to $\Gamma$ and $S^*_i = S_p$. In the particular case of
$G=U(1)$, and choosing the heat kernel action, $b_i(r) =
(\sqrt{2\pi\beta})^{-1}e^{-r^2/2\beta}$, one can express the partition
function in terms of $\theta$-functions
\cite{SpOl87},
\bea
\left\langle \prod_{p \in \Gamma} \chi_{S_p}(g) \right\rangle
&=& \left(\prod_{i\in\Gamma^*} e^{-(S^*_i)^2/2\beta}\right) \frac{
\sum_{h\in \Z} ~\exp \left\{-(2\beta)^{-1}N^*_0 h^2 + \beta^{-1} h
\sum_{i\in\Gamma} S^*_i \right\}}{\sum_{h\in \Z} ~\exp
\left\{-(2\beta)^{-1}N^*_0 h^2 \right\}} \nn\\ &=& \exp \left\{ -\frac
1 {2\beta} \left( \sum_{i\in\Gamma^*} (S^*_i)^2 - \frac 1
{N_0^*}\left( \sum_{i\in\Gamma^*} S^*_i\right)^2 \right)\right\}
\frac{\theta_3\left( \frac{2\beta}{N_0^*}; -\frac 1 {N_0^*}
\sum_{i\in\Gamma} S^*_i \right) }{\theta_3\left( \frac {2\beta}
{N^*_0}; 0 \right )}
\label{npoint}
\eea
For the (filled) Wilson loop, $\Gamma$ is a collection of adjacent
plaquettes with all $S_p = 1$. The overall factor in (\ref{npoint}) is
given by $\exp \left( - (2 \beta N_0^*)^{-1} A(\Gamma)\left(N^*_0 -
A(\Gamma)\right) \right)$, where $A(\Gamma)$ is the number of
plaquettes in $\Gamma$.  Note that this expression is symmetric under
interchanging the interior and exterior of $\Gamma$. In the
infinite volume limit, $N_0^* \rightarrow
\infty$, this reduces to the familiar area law (the $\theta$-function
factor in (\ref{npoint}) approaches 1). Using these techniques it is
possible to analyze the sector of the gauge theory in which there are
homologically non-trivial Wilson loops.

These few examples by no means exhaust the applications of our
formalism. A systematic study of its implications for confining gauge
theories in higher dimensions is in progress.


\begin{thebibliography}{1234567}
\newcommand{\bibi}[1]{\bibitem{#1}}
\newcommand{\authors}[1]{#1, }
\newcommand{\journal}[1]{#1}
\newcommand{\volume}[1]{{\bf #1}}
\newcommand{\myyear}[1]{(#1)}
\newcommand{\page}[1]{#1}
\newcommand{\mytitle}[1]{}
\newcommand{\keywords}[1]{}
\newcommand{\kw}[1]{}

\bibi{KrWa41}
\authors{H.A. Kramers and G.H. Wannier}
\journal{Phys. Rev.}
\volume{60} \myyear{1941} \page{252}.

\bibi{Sa80}
\authors{R. Savit}
\journal{Rev. Mod. Phys.}
\volume{52} \myyear{1980} \page{453}.

\bibi{Mu84}
\authors{J. Munkres}
\journal{ {\it Elements of Algebraic Topology}, Addison-Wesley, 
Menlo Park 1984}.

\bibi{DrWa82}
\authors{K. Dr\"uhl and H. Wagner}
\journal{Ann. Phys.}
\volume{141} \myyear{1982} \page{225}.

\bibi{FrKa80}
\authors{E. Fradkin and L.P. Kadanoff}
\journal{Nucl. Phys.}
\volume{170} \myyear{1980} \page{1}.

\bibi{CeKa71}
\authors{L.P. Kadanoff and H. Ceva}
\journal{Phys. Rev.}
\volume{B 3} \myyear{1971} \page{3918}.


\bibi{BuSh96}
\authors{A.I. Bugrij and V.N. Shadura}
\journal{JEPT} \volume{82} \myyear{1996} \page{552};
\journal{Phys.Rev.} \volume{B55} \myyear{1997} \page{1045};
\journal{Zh. Eksp. Teor. Fiz.} \volume{113} \myyear{1998} \page{240}.

\bibi{RaSe97}
\authors{M. Rakowski and S. Sen}
\journal{Lett. Math. Phys.}
\volume{42} \myyear{1997} \page{195}; \\
\authors{M. Rakowski}
\journal{Phys. Rev.}
\volume{D52} \myyear{1995} \page{354}.

\bibi{SpOl87}
\authors{J. Spanier and K.B. Oldham}
\journal{{\it An Atlas of Functions}, Hemisphere Publishing Corporation,
New York 1987}.

\end{thebibliography}
\end{document}